\documentclass[conference]{IEEEtran}
\IEEEoverridecommandlockouts
\usepackage{listings}
\usepackage{cite}
\usepackage{amsmath,amssymb,amsfonts}
\usepackage{algorithmic}
\usepackage{graphicx}
\usepackage{textcomp}
\usepackage{balance}
\usepackage{xargs}                      
\usepackage[pdftex,dvipsnames]{xcolor}  

\usepackage[hyphens]{url}
\usepackage{hyperref}
\hypersetup{breaklinks=true}

\usepackage[colorinlistoftodos,prependcaption,textsize=tiny]{todonotes}
\newcommandx{\unsure}[2][1=]{\todo[linecolor=red,backgroundcolor=red!25,bordercolor=red,#1]{#2}}
\newcommandx{\change}[2][1=]{\todo[linecolor=blue,backgroundcolor=blue!25,bordercolor=blue,#1]{#2}}
\newcommandx{\info}[2][1=]{\todo[linecolor=OliveGreen,backgroundcolor=OliveGreen!25,bordercolor=OliveGreen,#1]{#2}}
\newcommandx{\improvement}[2][1=]{\todo[linecolor=Plum,backgroundcolor=Plum!25,bordercolor=Plum,#1]{#2}}
\newcommandx{\thiswillnotshow}[2][1=]{\todo[disable,#1]{#2}}

\newcommand\archname{\textit{HPTMT}}

\def\BibTeX{{\rm B\kern-.05em{\sc i\kern-.025em b}\kern-.08em
    T\kern-.1667em\lower.7ex\hbox{E}\kern-.125emX}}
\begin{document}

\title{\archname: Operator-Based Architecture for Scalable High-Performance Data-Intensive Frameworks}

\author{
\IEEEauthorblockN{Supun Kamburugamuve\IEEEauthorrefmark{2},
Chathura Widanage\IEEEauthorrefmark{2},
Niranda Perera\IEEEauthorrefmark{1}, 
Vibhatha Abeykoon\IEEEauthorrefmark{3},
Ahmet Uyar\IEEEauthorrefmark{2},\\
Thejaka Amila Kanewala\IEEEauthorrefmark{3},
Gregor von Laszewski \IEEEauthorrefmark{2}
and Geoffrey Fox\IEEEauthorrefmark{4}\IEEEauthorrefmark{2}
}

\IEEEauthorblockA{\IEEEauthorrefmark{1}Luddy School of Informatics, Computing and Engineering, Bloomington, IN 47408, USA\\
dnperera@iu.edu}
\IEEEauthorblockA{\IEEEauthorrefmark{2}Digital Science Center, Bloomington, IN 47408, USA\\
\{skamburu, cdwidana, auyar\}@iu.edu, laszewski@gmail.com}
\IEEEauthorblockA{\IEEEauthorrefmark{3}Indiana University Alumni, IN 47408, USA\\
\{vibhatha,thejaka.amila\}@gmail.com}
\IEEEauthorblockA{\IEEEauthorrefmark{4}From August 2021, Biocomplexity Institute \& Initiative and Computer Science Dept., University of Virginia\\
gcfexchange@gmail.com}

}

\maketitle

\begin{abstract}
Data-intensive applications impact many domains, and their steadily increasing size and complexity demands high-performance, highly usable environments. We integrate a set of ideas developed in various data science and data engineering frameworks. They employ a set of operators on  specific data abstractions that include vectors, matrices, tensors, graphs, and tables. Our key concepts are inspired from systems like MPI, HPF (High-Performance Fortran), NumPy, Pandas, Spark, Modin, PyTorch, TensorFlow, RAPIDS(NVIDIA), and OneAPI (Intel). Further, it is crucial to support different languages in everyday use in the Big Data arena, including Python, R, C++, and Java. We note the importance of Apache Arrow and Parquet for enabling language agnostic high performance and interoperability. In this paper, we propose \textit{High-Performance Tensors, Matrices and Tables} (\archname{}), an operator-based architecture for data-intensive applications, and identify the fundamental principles needed for performance and usability success. We illustrate these principles by a discussion of examples using our software environments, Cylon and Twister2 that embody \archname{}.
\end{abstract}


\section{Introduction}

Data-intensive applications have evolved rapidly over the last two decades, and are now being widely used in industry and scientific research. Large-scale data-intensive applications became mainstream with the rise of the \textit{map-reduce programming paradigm}. However, the data engineering community has come a long way in integrating the idea of \textit{map-reduce}. There is a broader understanding on the data engineering application classes, and specialized frameworks that serve them. Modern systems can crunch data and learn from them using sophisticated algorithms that even make use of custom hardware solutions. We have seen different programming models and APIs being developed to make it easier to program data-intensive applications. 

Due to the diverse nature of data-intensive applications, it is hard for one framework to support all classes of problems efficiently. For example, we may need to load data, curate them, utilize machine learning algorithms, conduct post-processing, and perform visualizations, all as parts of single application pipeline. Such an integration is done either as a custom-developed single program or by developing the pieces separately and combining them into a data-intensive application workflow. The complexities are vast, and having interoperable systems enhances usability significantly.

The various application classes require tailored abstract concepts. Vectors, matrices, tables, graphs, and tensors are widely used examples in data-intensive computations. For applications to benefit from these abstractions efficiently, operations around them must be implemented to provide solutions for general-purpose and problem-specific domains. Among these operations,  matrix multiplication, vector addition, and table joins are some standard operations. We can represent these abstract objects in the main memory via data structures; vectors, matrices, and tensors are shown as arrays, graphs are represented either using matrices or as edge lists, and tables are configured as a set of columns or rows. One example is Apache Spark~\cite{spark2010}, which has a table abstraction originating from relational algebra. Deep learning systems such as PyTorch~\cite{pytorch} are based on tensor abstractions originating from linear algebra. 

When developing applications with modern frameworks, we often resort to data abstractions and their operators that may not fit the original problem because required data abstractions are not supported.
This is due to lack of cohesiveness amongst systems and in-turn leaves a lot of performance on the table, even though they are capable of doing a top-notch job at their intended purpose. Ideally we would like different data abstractions and operators to work together to solve problems. In order to address this issue, we will analyze the fundamental designs of these systems while focusing on the elementary operators they provide and how they can work hand-in-hand. 

We introduce the \archname(High Performance Tensors, Matrices and Tables) architecture in this paper, which defines an operator-centric interoperable design for data-intensive applications. The authors have developed two frameworks called Cylon~\cite{widanage2020high} and Twister2~\cite{twister2} aimed at developing data-intensive applications. We will take these as examples to showcase the importance of designing operators and how the various systems implementing them can work together according to \archname{} architecture.

The rest of the paper is organized as follows. Section~\ref{sec:arch} gives a high-level introduction and motivation for \archname{}. Section~\ref{sec:arrays} describes arrays and distributed operators around them, while Section~\ref{sec:tables} focuses on tables. The next two sections,~\ref{sec:models} and \ref{sec:exec}, talk about programming models and execution models around these data structures. Section~\ref{sec:archi} presents the operator-based architecture, and Section~\ref{sec:frame} discusses how this architecture is realized in Twister2 and Cylon. 

\section{HPTMT Architecture}
\label{sec:arch}

One of the most successful approaches to parallel computing is based on the use of runtime libraries of well-implemented parallel operations. This was for example a key part of High-Performance Fortran HPF~\cite{dongarra2003sourcebook} and related parallel environments (HPJava~\cite{carpenter1998hpjava}, HPC++~\cite{johnson1997hpc}, Chapel~\cite{chamberlain2007parallel}, Fortress~\cite{allen2005fortress}, X10~\cite{charles2005x10}, Habanero-Java~\cite{imam2014habanero}). Such systems had limited success; maybe because the HPC community did not define sufficient operators to cover the sophisticated computational science simulations largely targeted by those languages with typically sparse or dense matrix operators. However data-intensive applications have used similar ideas with striking success. 

Perhaps, the most dramatic event was the introduction of MapReduce~\cite{dean2008mapreduce} some 15 years ago, and its implementation in Hadoop which enabled parallel databases as in Apache Hive. MapReduce adds Group-By and key-value pairs to the Reduce operation common in the simulation applications of the previous HPF family. The powerful yet simple MapReduce operation was expanded in Big Data systems especially through the operators of Databases (union, join, etc.), Pandas (we identify 244 dataframe operators out of 4782 Pandas methods), and Spark, Flink~\cite{flink2015}, Twister2 (~70). Deep Learning environments such as PyTorch, TensorFlow~\cite{abadi2016tensorflow} (with Keras~\cite{gulli2017deep}) added further (over 700) operators to build deep learning components and execution. 

The powerful array libraries of Numpy~\cite{van2011numpy} are built on a large (at least 1085) set of array operations used in the original scientific simulation applications. Oversimplifying HPF as built around matrix or array operators, we suggest today that the natural approach is HPTMT or High-Performance Tables, Matrices, and Tensors. Operator based methods are not just used to support parallelism but have several other useful capabilities

\begin{itemize}
    \item Allow interpreted languages to be efficient as overhead is amortized over the execution of a (typically large) operation
    \item Support mixed language environments where invoking language (e.g. Python) is distinct from the language that implements the operator (e.g. C++)
    \item Support proxy models where user programs in an environment that runs not just in a different language but also on a different computing system from the executing operators. This includes the important case where the execution system includes GPUs and other accelerators.
    \item Support excellent performance even in non-parallel environments. This is the case for Numpy and Pandas operators.
\end{itemize}

\archname{} is supported by many libraries including those like ScaLAPACK~\cite{blackford1997scalapack} (320 functions with a factor 4 more counting different precisions) and its follow-ons such as Intel’s MKL (oneAPI and Data Parallel C++)~\cite{intel_one_api} originally motivated by HPF simulation goals but equally important for Big Data. The NVIDIA RAPIDS~\cite{rapids} project is building a GPU library covering much of \archname{} requirements as is our Cylon project for CPUs. Modin~\cite{petersohn2020towards} is using Dask~\cite{rocklin2015dask} and Ray~\cite{moritz2018ray} for parallel Pandas operators.

Recently Apache Arrow~\cite{apache_arrow} and Parquet~\cite{parquet} have been developed providing important tools supporting \archname{}. They provide efficient language agnostic column storage for Tables and Tensors that allows vectorization for efficiency and performance. Note that distributed parallel computing performance is typically achieved by decomposing the rows of a table across multiple processors. Then within a processor, columns can be vectorized. This of course requires large amount of data so that each processor has a big-enough workset to processes efficiently. It is a well-established principle that the problem needs to be large enough for the success of parallel computing~\cite{fox2014parallel}, which the latest Big Data trends also follow. Note that in scientific computing, the most effective parallel algorithms use block (i.e. row and column) decompositions to minimize communication/compute ratios. Such block decompositions can be used in Big Data~\cite{huai2014major} (i.e. table data structures), but could be less natural due to the heterogeneous data within it. 

For Big Data problems, individual operators are sufficiently computationally intensive that one can consider the basic job components as parallel operator invocations. Any given problem typically involves the composition of multiple operators into an analytics pipeline or more complex topology. Each node of the workflow may run in parallel. This can be efficiently and elegantly implemented using workflow (Parsl~\cite{babuji2018parsl}, Swift~\cite{wilde2011swift}, Pegasus~\cite{deelman2015pegasus}, Argo~\cite{argo}, Kubeflow~\cite{kubeflow}, Kubernetes~\cite{kubernetes}) or dataflow (Spark, Flink, Twister2) preserving the parallelism of HPTMT.

\section{Arrays}
\label{sec:arrays}

Arrays are a fundamental data structure of scientific computing, machine learning, and deep learning applications because they represent vectors, matrices, and tensors. An array consists of elements from the same data type, and an index can address each value. An array is stored in contiguous memory of a computer. The size of an element and the index can efficiently compute the memory location of an array element. A single value array of a type is equivalent to a scalar of that type. As such, arrays can represent all primitive types. Variable width data types such as Strings would require a composite arrays that represent data and offsets/ strides. 

\subsection{Vectors and Matrices}

We can represent a vector directly using an array. A matrix is a 2-dimensional grid, and each value can be addressed using a row index and a column index. We need to use 2-dimensional arrays or 1-dimensional arrays to represent a matrix. We can store a matrix in row-major format or column-major format. In row-major format, the values of a row are in the contiguous memory of the array. In contrast, the column values are found in the array's contiguous memory with the column-major format.. These formats are designed to match the access patterns of matrices when doing calculations. 

\subsubsection{Sparse Matrices}

Sparse matrices are defined as matrices where the majority of elements are zero. To store these as regular dense matrices wastes both memory and CPU cycles during computations. As an alternative, there are efficient layouts such as Compressed Sparse Column (CSC), Compressed Sparse Row (CSR), and Doubly Compressed Sparse Column (DCSC) for sparse matrices. All these formats use arrays to store the matrix in memory.

\subsection{Tensor}

A tensor can be interpreted of as a more generic view on a collection scalars or vectors or both to represent a mathematical model or data structure. A matrix is also a tensor by definition, and it is the most generic abstraction for mathematical computations. Similar to matrices, tensors can be stored according to the access patterns using arrays.

\subsection{Operations}

Distributed operations around arrays are defined in the MPI (Message Passing Interface)~\cite{MPI-3.0_2012} standard as collective operations. The table~\ref{tab:array_op} describes some of these operations, which are derived from common communication patterns involved when dealing with vectors and matrices in the form of arrays. They are optimized to work on thousands of computers using data transfer algorithms~\cite{wickramasinghe2016survey} that can minimize the latency and utilize the available network bandwidth to the fullest.

\begin{table}[]
\caption{Array-based distributed operations as specified by MPI} 
\label{tab:array_op}

\centering
\begin{tabular}{|p{2cm}|p{6cm}|}
\hline
\textbf{Operation} & \textbf{Description}                                         \\ \hline
Broadcast          & Broadcast an array from one process to many other processes. \\ \hline
Gather/AllGather & Collects arrays from different processes and creates a larger array in a single process or many processes. \\ \hline
Scatter/AllToAll   & Redistributes the parts of an array to different processes.  \\ \hline
Reduce/AllReduce & Element-wise reduction of arrays. Popular operations include SUM, MIN, MAX, PROD.                          \\ \hline
\end{tabular}
\end{table}

\section{Tables}
\label{sec:tables}

A table is an ordered arrangement of data into a 2-dimensional rectangular grid with rows and columns. A single column of a table has data of the same type, while different columns can have different data types (heterogeneous data). 

\subsection{Tables in Memory}

We can store table data in main memory using a simple technique such as a list of records or compact formats that keep the values in contiguous memory. Compact memory layouts can store tables similar to row-major and column-major representations in matrices. Having a list of records can lead to inefficient use of memory and degrade the performance of applications due to cache misses, TLB (Translation Lookaside Buffer) being misused, and serialization/deserialization costs.  

In a column-major representation, the table columns are stored in contiguous memory. Table representations are complex compared to matrices because they can have variable-length data. In such cases, length information needs to be stored along with the columns. Thus, sparsity can be embedded or integrated into separate arrays. Also, information about \textit{null} values needs to be stored in a table. One benefit of a column-major representation is that the values of the column are of the same data type. In a row-major definition, the table rows can be stored in contiguous memory. This means that different types of values are stored in contiguous memory with different bit widths. So we need to keep track of lengths and NULL values.

\subsection{Operations}

We can use row-based partitioning, column-based partitioning, or a hybrid version to divide table data into multiple processes. Most of the time, data processing systems work on tables distributed with row-based partitioning. Relational algebra defines five base operations around tables described in Table~\ref{tab:table_found_op}. Table~\ref{tab:table_aux_op} lists some commonly used auxiliary operations around tables. Popular table abstractions like Pandas~\cite{mckinney2011pandas} extend these to hundreds of operators. 

\begin{figure}[!h]
\centering
\includegraphics[width=0.4\textwidth]{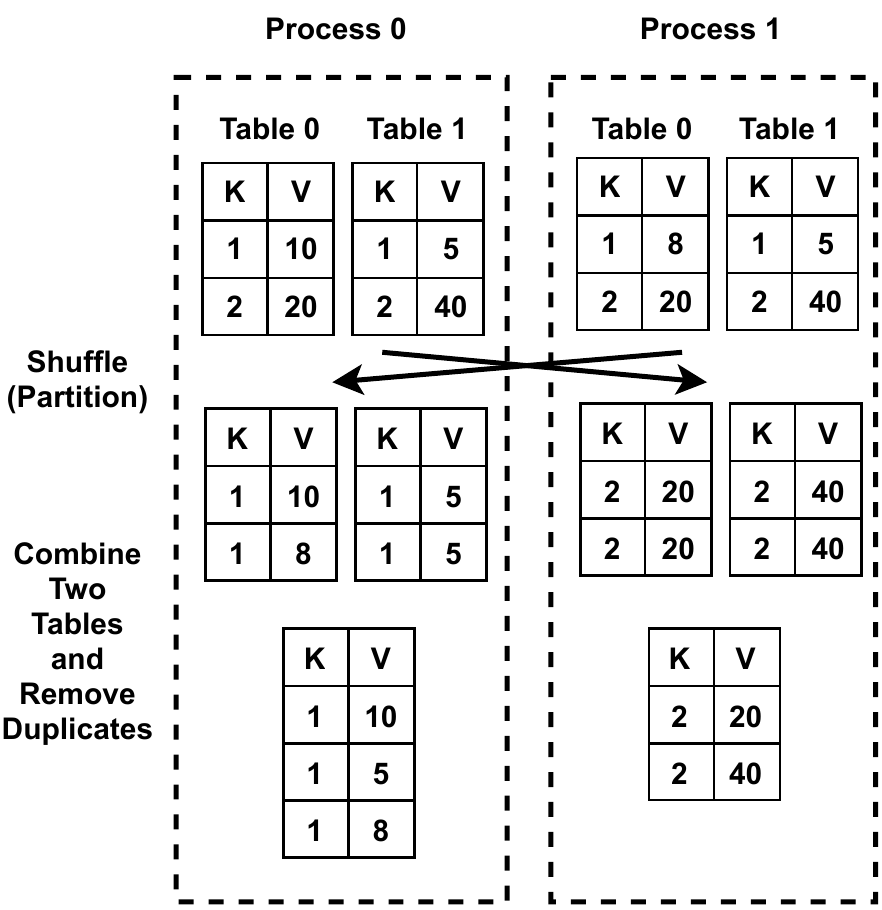}
\caption{Union of two tables distributed in two processes}
\label{fig:union}
\end{figure}

Figure~\ref{fig:union} shows two tables in two processes and a distributed union operation that removes duplicates. This operation needs to redistribute the records of the tables so that the identical records go to the same table. Such redistribution is common in table-based distributed operations and is commonly referred to as \textit{shuffle} operation. 

\subsubsection{Shuffle}

\begin{figure}[!h]
\centering
\includegraphics[width=0.35\textwidth]{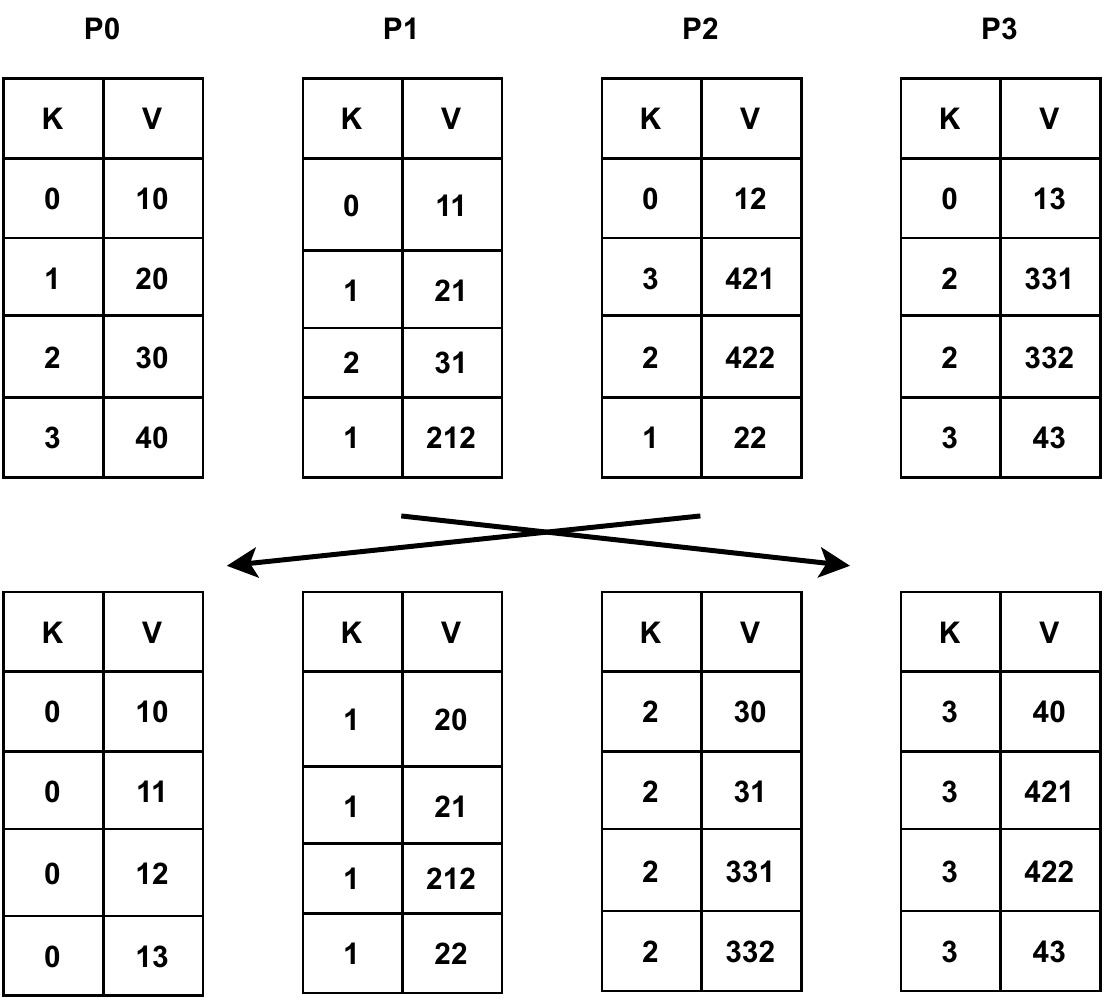}
\caption{Shuffle of 4 tables in 4 processes}
\label{fig:shuffle}
\end{figure}

Figure~\ref{fig:shuffle} displays a  shuffle operation applied to four tables in four processes. The shuffle is done on attribute \textit{K}. Shuffle is similar to the array \textit{AllToAll} operation, which is equivalent to every process scattering values to other processes. Shuffle is similar in that it scatters records of a table to every other process. What makes these two operations different are the data structure, its representation in memory, and how we select which values are scattered to which processes. In \textit{AllToAll}, scatter occurs by a range of indexes. In tables, the shuffle takes place based on a set of column values.

Large-scale data operations require careful use of memory and optimizations to scale to a large number of cores~\cite{barthels2017distributed}. Because of the nature of these data structures like tables and arrays, we can use one structure to represent another. For example, we can have a table to represent a matrix. Also, we can have a set of arrays to represent a table. Therefore, we can sometimes use a programming API (data structures and operators) developed for a specific set of problems, to solve problems in unrelated/ unintended domains. In practice, this leads to unnecessary inefficiencies in execution. For instance, say we have a table abstraction with GroupBy and aggregate operations implemented. Now we want to get AllReduce-sum semantics of a column in this table. We can do so by assigning a common key to each value in columns and doing a GroupBy on the key, followed by an aggregate operation. However, this is not an efficient method because it uses an additional column and a shuffle operation, which is more costly than the communication required for an AllReduce operation. 

Arrays and tables have their own distributed operations. As seen in previous sections, the distributed operations on tables originate from relational algebra, and those on arrays derived to support linear algebra. 

\begin{table}[!h]
\caption{Fundamental table operations} 
\label{tab:table_found_op}
\centering
\begin{tabular}{|p{1.5cm}|p{6.5cm}|}
\hline
\textbf{Operator} & \textbf{Description}                                                   \\ \hline
Select            & Filters out some records based on the \& value of one or more columns. \\ \hline
Project           & Creates a different view of the table by dropping some of the columns. \\ \hline
Union             & Applicable on two tables having similar schemas to keep all the records from both tables and remove duplicates. \\ \hline
Cartesian Product & Applicable on two tables having similar schemas to keep only the records that are present in both tables.       \\ \hline
Difference        & Retains all the records of the first table, while removing the matching records present in the second table.    \\ \hline
\end{tabular}
\end{table}

\begin{table}[!h]
\caption{Auxiliary table operations} 
\label{tab:table_aux_op}

\centering
\begin{tabular}{|p{1.5cm}|p{6.5cm}|}
\hline
\textbf{Operator} & \textbf{Description}                                                                                               \\ \hline
Intersect         & Applicable on two tables having similar schemas to keep only the records that are present in both tables.          \\ \hline
Join              & Combines two tables based on the values of columns. Includes variations Left, Right, Full, Outer, and Inner joins. \\ \hline
OrderBy          & Sorts the records of the table based on a specified column.                                                        \\ \hline
Aggregate & Performs a calculation on a set of values (records) and outputs a single value (Record). Aggregations include summation and multiplication. \\ \hline
GroupBy           & Groups the data using the given columns; GroupBy is usually followed by aggregate operations.                      \\ \hline
\end{tabular}
\end{table}

\section{Programming Models and Operators}
\label{sec:models}

Data-intensive applications use both implicit and explicit parallel programming models. In an explicit model, the user is aware of the parallel nature of the program, writes the application according to a local view, and synchronizes the data distributed in multiple computers using distributed operations. MPI-based programming is the most popular explicit model. Most data-intensive applications use an implicit parallel model with a distributed data abstraction. These models have similarities to partitioned global address space models~\cite{zheng2014upc, chamberlain2007parallel}. 

\subsection{Local Data Model}

In this model, the user programs a parallel process or task. Here, users only deal with the local data, and when they need to synchronize it with other processes, they invoke a distributed operator.  

\begin{figure}[!h]
\begin{lstlisting}[language=Java]
// every process loads its own data
LocalData A = readFiles()  
// apply local operators
LocalData B = A.filter(Function filter) 
// sort is a distributed operator, the data 
// of B in the parallel processes will be sorted
LocalData C = sort(B)
// save C to disk
C.save()     	
\end{lstlisting}
\caption{Eager execution}
\label{fig:eager}
\end{figure}

\subsection{Global Data Model}

In the distributed data API, the user defines an abstract object that acts as a global view for the data distributed across the cluster. This object represents a dataset such as a table or an array. The user applies operations to this global data that produces more distributed data objects. This is an implicit parallel programming model that has been present for a while in various forms under the general umbrella of partitioned global address space (PGAS) programming model and is used by data-intensive frameworks extensively. Depending on the amount of data processed, we can have an eager model or a dataflow model using external storage for computations. 

\subsubsection{Eager Model}

With an eager model, the operators work on in-memory data and can be executed immediately. Combining SPMD (Single Program Multiple Data)-style user code and distributed data-based API for the data structures, we can create powerful and efficient APIs for data-intensive applications.
We depict the code in Figure \ref{fig:in-memory}. We assume ‘A’ is representing a partitioned table in multiple computers. Now ‘B’ and ‘C’ are also partitioned tables.  

\begin{figure}[!h]
\begin{lstlisting}[language=Java]
// load the data as partitions on 
// multiple processes
DistributedData A = readFiles() 
// user supplies a filter function
DistributedData B = A.filter(Function filter) 
// sort is a distributed operator
// that requires network communication
DistributedData C = B.sort()    
C.save()    	   
\end{lstlisting}
\caption{In-memory API}
\label{fig:in-memory}
\end{figure}

This code will run as SPMD code, but the data structures and operators can reduce the programmer's burden by providing a global data structure. The above code uses the distributed memory model with no threads. 

\subsubsection{DataFlow Model}

Data-intensive applications constantly work with datasets that do not fit into the available random access memory of computing clusters. To work with large datasets, we need to support streaming computations that utilize external storage. The previous API with eager execution is not suitable here, and we can illustrate this with an example (see Figure \ref{fig:eager}). Here we need to save data multiple times to the disk to execute the program.

\begin{figure}[!h]
\begin{lstlisting}[language=Java]
// A is a large dataset, so we need to 
// store it in the disk
DistributedData A = readFiles()  
// B is still large, so we need to store 
// B in the disk
DistributedData B = A.filter(Function filter) 
// to sort B, we need to use disk and 
// then store C in disk
DistributedData C = B.sort() 
// save C to disk
C.save()     	
\end{lstlisting}
\caption{Eager execution}
\label{fig:eager}
\end{figure}

We can avoid such extensive use of external storage by executing the above program as a single graph with data streaming through it piece by piece. This is called the \textit{dataflow model}. In this model, computations are data-driven (event-driven). The sources produce data, and this data activates the next connected computation. The middle nodes can produce more data until they reach a sink node without an output. The links represent distributed operators that carry data between nodes that can be within the same process or in different processes across machines. There are usually constraints applied to nodes that force them to be scheduled into computers in separate ways depending on the framework and the application. 

The functionality of links depends on the operators and the data abstractions they represent. Each user-defined function runs on its own program scope without access to any state about other tasks. The only way to communicate between dataflow nodes is by messaging, as they can run in various places. This model is adopted by popular data processing engines such as Spark and Flink~\cite{flink2015}.

\section{Execution}
\label{sec:exec}

Many parallel programs are instances of a single program running multiple instances working on different data, or multiple programs doing the same. This is called single program multiple data (SPMD) and multiple programs multiple data (MPMD) style parallel programs. SPMD programs are popular in batch programs where the same task instances (program) are executed in parallel processes simultaneously. For example, Hadoop map-reduce~\cite{white2012hadoop} programs are executed as SPMD programs where the map tasks are executed first, and then the reduce tasks are executed in parallel. MPMD programs are used in streaming applications and pipeline parallel batch programs. In an MPMD program, different tasks run on separate parallel processes. Apache Storm~\cite{toshniwal2014storm} and Flink run streaming programs in this style.

\begin{figure}[h]
\centering
\includegraphics[width=0.45\textwidth]{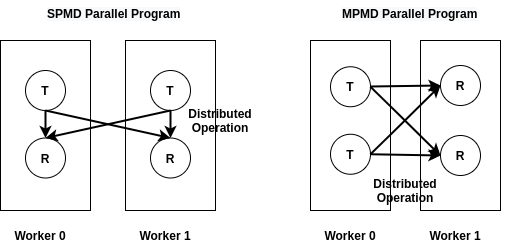}
\caption{SPMD and MPMD Programs}
\label{fig:sync-1}
\end{figure}

\subsection{Memory Models}

We can further divide parallel programs according to how parallel tasks share data. Here we have shared memory, distributed memory, and hybrid memory execution models. In a shared memory model, parallel instances of the program share the same memory address space. This model only allows parallel programs to run in a single computer that has many CPU cores. In a distributed memory model, every instance of the parallel program is executed on an isolated memory such as its own process. They can only access the data with processes using messaging. In hybrid memory model, some instances of parallel processes are run in the shared memory model. These groups of parallel instances need messaging to share data with other such groups.

\subsection{Loosely Synchronous and Asynchronous Execution}

We can also execute parallel applications in a loosely synchronous way or asynchronously. The loosely synchronous execution as shown in Figure~\ref{fig:sync-1} assumes all the tasks are executing concurrently, and the processes synchronize with each other by exchanging messages at certain points. The sections of code between communication synchronizations execute independently from other parallel processes.

\begin{figure}[h]
\centering
\includegraphics[width=0.4\textwidth]{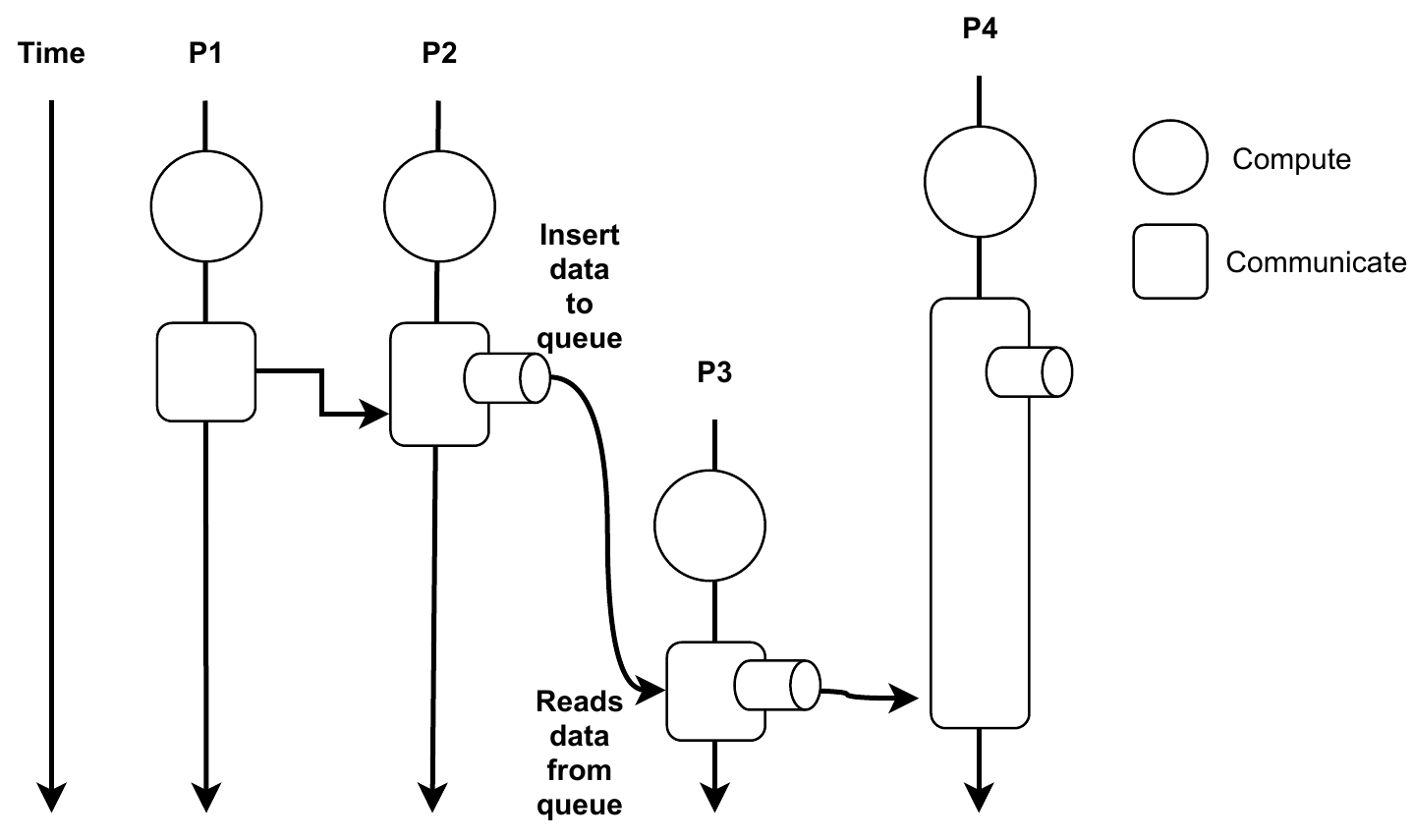}
\caption{Asynchronous Execution}
\label{fig:async-1}
\end{figure}

In asynchronous execution, as shown in Figure~\ref{fig:async-1}, the tasks are decoupled in time. When a task sends a message, we can assume it is stored in a queue. Once the receiving task is ready, it picks up the message. This allows more flexible execution of the tasks independent of the programming model, but often needs a central server as a facilitator that notifies the receivers about pending messages.  Also, storing a message, notifying the receiver about the message, and picking it up creates a middle step that can reduce the performance. Asynchronous execution has parallels to the preemptive scheduling we see in operating systems where it tries to simultaneously progress multiple programs to increase the responsiveness. 

The asynchronous execution described here is an implementation detail because the synchronization comes from the operator. For example, in a distributed join operation, a single process requires data from all the other processes. So whether we insert messages into a queue or not, the program cannot continue until the operation is completed by receiving messages from other processes. 

\begin{figure}[h]
\centering
\includegraphics[width=0.4\textwidth]{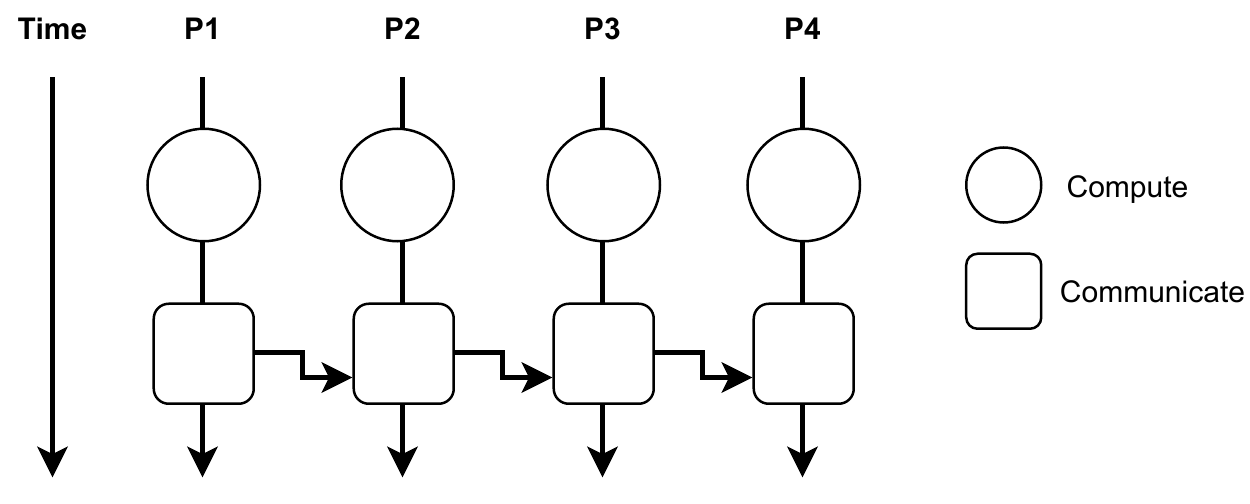}
\caption{Loosely Synchronous Execution}
\label{fig:sync-1}

\end{figure}

\section{\archname{} Architecture Principals}
\label{sec:archi}

\begin{figure}[h]
\centering
\includegraphics[width=0.25\textwidth]{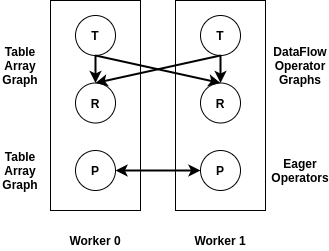}
\caption{Operator Architecture for Data-Intensive Applications}
\label{fig:spos}
\end{figure}

We define \archname{} as an architecture where any combination of loosely synchronous operators built around different data abstractions working together to develop data-intensive applications. A high-level view of the architecture is shown in Figure~\ref{fig:spos}, where dataflow operators and eager operators work together in a single parallel program. 

\begin{figure}[h]
\centering
\includegraphics[width=0.35\textwidth]{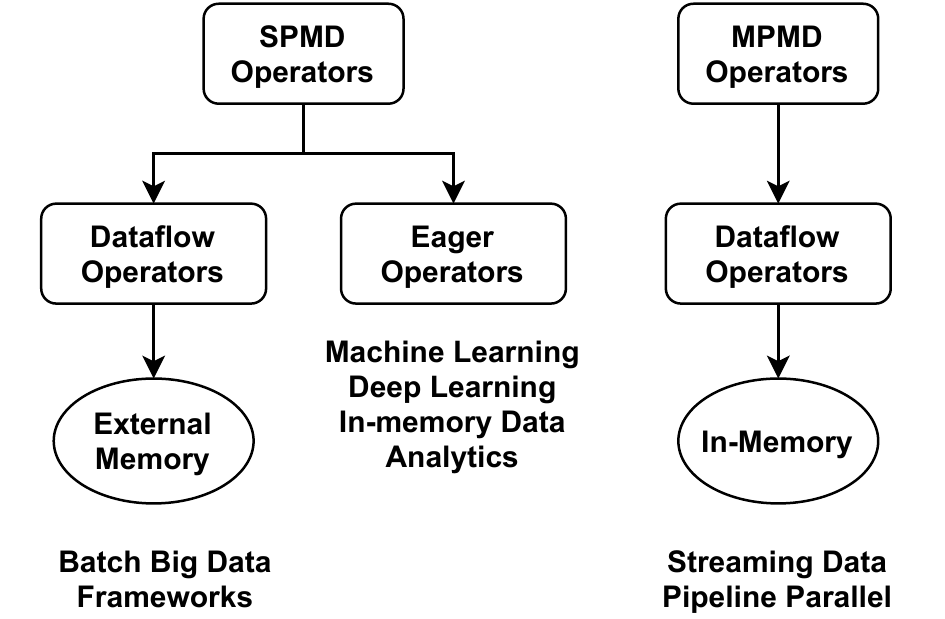}
\caption{Operators with SPMD and MPMD programs}
\label{fig:operators}
\end{figure}

In general, we can categorize  operators depending on whether they are designed to run on MPMD or SPMD executions. This is shown in Figure~\ref{fig:operators}. Here SPMD-style programs can use dataflow-style operators for programs that demand external memory, or eager operators for applications that run in memory. MPMD-style operators tend to be dataflow operators because they are used in streaming and pipeline parallel programs.  

\subsection{Dataflow and Eager Operators}

In general, a dataflow operator needs to take input piece-by-piece and produce output the same way. It can also take input at once and produce output individually, or take input piece-by-piece and produce output as a single object as well. This means it needs to be a non-blocking operator. Furthermore, the operators need to determine when to terminate by using a termination algorithm because inputs are not coordinated. If it is a streaming application, the operator may not terminate and continue consuming and producing data. Dataflow operators can use external storage to keep the intermediate data in order to do operations that cannot fit into the memory.

Eager operators work on in-memory data by taking input data at once and producing output all at once. This is the approach taken by operators in MPI. They can be deterministic in terms of execution because only one data input is given and one output is produced. 

\subsection{SPMD and MPMD Operators}

In an SPMD-style program, the same processes participate in the operators as data producers and receivers. This can simplify the operator interfaces. For example, the collective operators in MPI standard are implemented in this fashion.

In an MPMD-style program, the operators can have data producers and receivers in different processes. This is a more general form of an operator, as it can represent SPMD-style operators as well. Twister:Net~\cite{t2-net} is one such MPMD-style operator library. Whether it is SPMD or MPMD, we can have eager style operators or dataflow operators. Streaming systems are where we primarily see MPMD-style operators.

\subsection{ Operator Principles}

Whether it is a dataflow operator or an eager operator, \archname{} architecture identifies several design principles for them to work together in different environments. 

\begin{itemize}
    \item Multiple data abstractions and operators - Discourage the use of data structures and operators suitable for one class of problems to be used in another class of problems. i.e. Do not use table operators for a problem that needs arrays. 
    \item Efficient Loosely Synchronous Execution - In an asynchronous framework, operators and the scheduler are coupled. In such situations we may need to develop operators specifically targeting the framework, which is contrary to the \archname{} goals. 
    \item Independence of the parallel execution environment - A parallel environment manages the processes and various resources required by operators, such as the network. If the implementation of operators is coupled to the execution environment, we can only use the operators specifically designed for it. We see this design in MPI-based operators where collectives are coupled to the MPI implementation's parallel process management. Frameworks such as UCX~\cite{shamis2015ucx} are in the process of developing MPI-equivalent collective operators for arrays without process management. In other words we should be able to bootstrap operator implementation on various parallel environments. 
    \item Same operator on different hardware - The same operator can be implemented on GPUs, CPUs or FPGA (Field Programmable Gate Arrays). Also, they should be able to use different networking technologies such as Ethernet and InfiniBand.
\end{itemize}

In some situations, we can get around certain design principles and make operators work together. For example, we can use MPI primitive based operators on different data abstractions as long as we are running within an MPI execution environment. So it partially satisfies the \archname{} requirements.

\begin{figure}[h]
\centering
\includegraphics[width=0.4\textwidth]{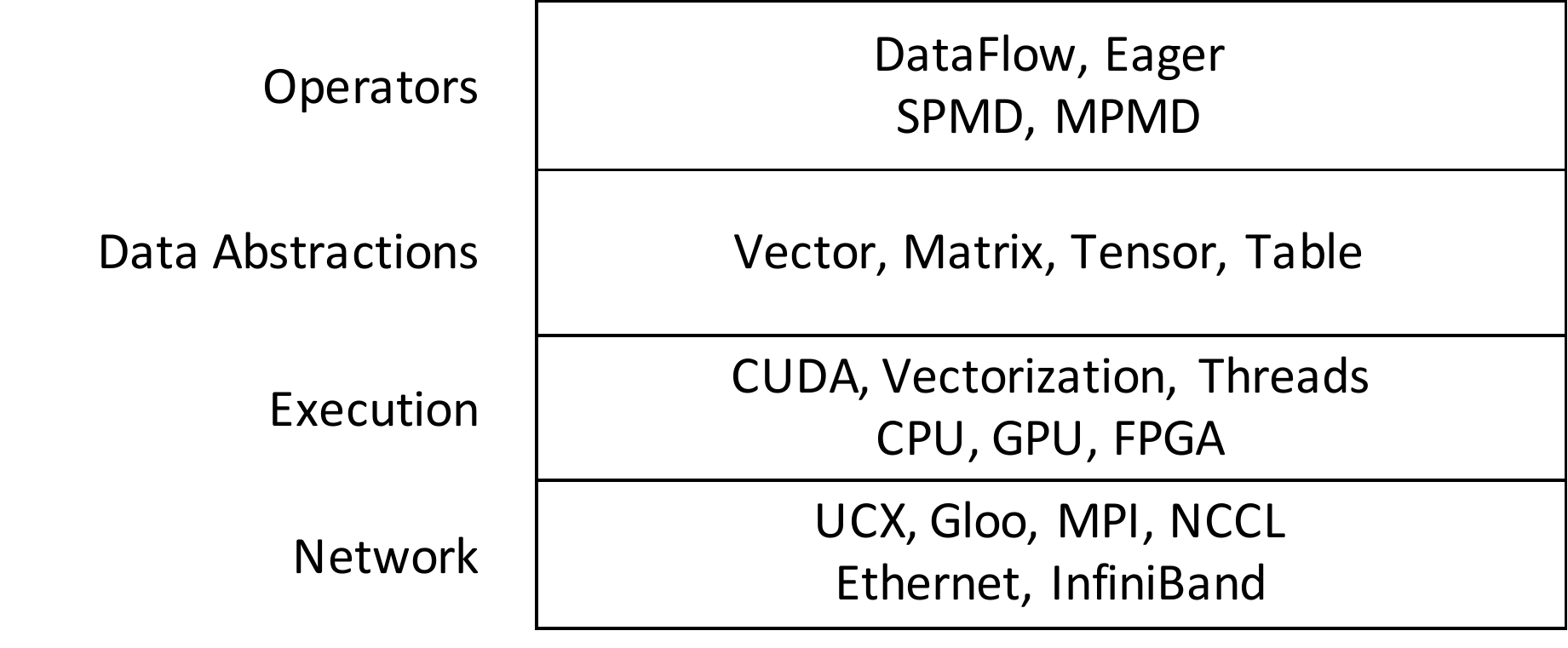}
\caption{Distributed operator implementation layers}
\label{fig:operator_layers}
\end{figure}

The general architecture of distributed operators is shown in Figure~\ref{fig:operator_layers}. At the bottom are the networking hardware and various software libraries that abstract out them. Then we have different execution units such as CPUs and GPUs. These are also abstracted by various programming APIs. On top of these we have our data structures in the memory defined according to various formats. We define the distributed and local operators on these data structures using the various execution and networking hardware.

\subsection{Workflows}

A workflow is a sequence of tasks connected by data dependencies. Given such a set of tasks, a workflow system orchestrates the execution of the tasks in the available resources preserving the data dependencies. Workflow systems for scientific computing applications~\cite{yu2005taxonomy} have been around for some time, with prominent scientific workflow systems such as Kepler~\cite{ludascher2006scientific} and Pegasus~\cite{deelman2015pegasus} leading the way. In addition, there are data-intensive application-specific workflow systems such as Kubeflow and Apache Airflow.

Workflow systems provide mechanisms for specific applications using domain-specific languages (DSLs) as well as using graphical user interfaces. Furthermore, we can use general-purpose programming languages to specify workflows as seen in Python-based systems such as Parsl~\cite{babuji2018parsl} and Ray-Project~\cite{}.

It is important to make the distinction between tasks of a parallel program and tasks of a workflow. A workflow task system is usually an application such as a machine learning algorithm. This algorithm may need to run on multiple computers, and it might run internally as a set of tasks. These internal tasks to the machine learning algorithm are fine-grained and usually developed using the programming methods we described earlier. It is not efficient to use workflows for finer-grained tasks because of the central coordination.

\subsubsection{Remote Execution}

Remote execution is a form of workflow adopted by current data systems. With this model, a data-intensive program is created at a central server and submitted to the cluster to execute. We believe having a clean separation between programs and workflow is important for the inseparability of frameworks. Mixing both can hinder development and make it difficult for programmers to think about applications.

\subsection{Separation of Concerns}

Separation of concerns is a design principle that states we should separate a program into distinct sections that address specific concerns. For example, from a parallel computing perspective, running computations on a cluster is a separate concern better addressed by a workflow engine designed precisely for that task. Developing and running a parallel application is another task that should be handled by frameworks suited for those tasks. The remote execution methods adopted by current programs combine these two aspects into a single program. 

\begin{figure}[h]
\centering
\includegraphics[width=0.4\textwidth]{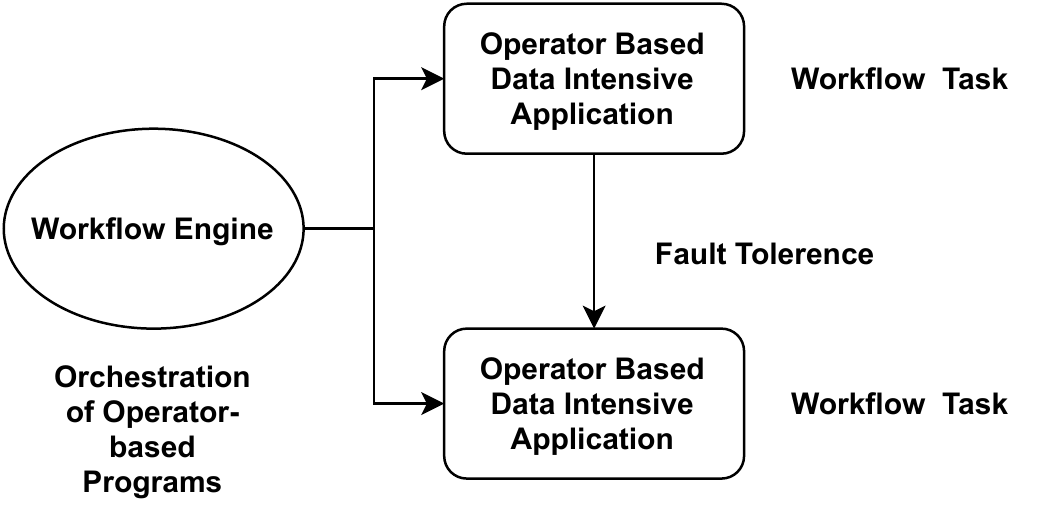}
\caption{Workflow for orchestrating operator-based data-intensive applications}
\label{fig:workflow}
\end{figure}

Going by these concerns, we propose operator-based data-intensive applications orchestrated by a workflow engine as the overall architecture of data intensive applications as shown in Figure~\ref{fig:workflow}.

We can combine distributed operator-based parallel programs with workflow engines to create rich data-intensive applications. Aspects such as fault handling can be moved to the workflow level to keep a balance between performance and fault tolerance overheads.

\subsection{Fault Tolerance}

Handling faults is an important aspect of large-scale data-intensive applications. Modern hardware is becoming increasingly reliable, and the chances of hardware failure during a computation are decreasing. But for applications that execute over a longer duration, handling hardware and software failure is still important. Streaming applications and long-running batch applications are some examples.  

Handling faults at the operator level is inefficient as it adds additional synchronization to communication steps. Because of this, we can always handle the faults outside of the operator code. If an operator fails, we can go back to the previous state before starting the operator. But operators need to detect failures and notify the applications to handle them gracefully. 

\section{Frameworks}
\label{sec:frame}

Let us look at Twister2 and Cylon, which are frameworks with DataFlow and eager operator APIs for data-intensive applications. We take these as examples for the proposed \archname{} architecture in combination with array based operators of MPI. 

\subsection{Twister2}

Twister2~\cite{twister2} is a data analytics framework for both batch and stream processing. The goal of Twister2 is to provide users with performance comparable to HPC systems while exposing a user-friendly dataflow abstraction for application development. TSet~\cite{wickramasinghe2019twister2} is the distributed data abstraction of Twister2. Twister2 provides a table abstraction and array abstraction with DataFlow operators. This allows Twister2 to compute using external memory efficiently for problems that do not fit into the memory. Twister2 operators are implemented as MPMD and can support both streaming and batch applications.

Local and distributed operations can transform or combine TSets to produce new TSets with different schemas. The simplest model of TSet is modeling a distributed primitive array (series). At the same time, this can be extended to represent a table by making every element of an array a composite object, as shown in Figure \ref{fig:tset_mpi}. Although the worker level parallelism of Twister2 is set to four at line 5, TSet level parallelism can even be a different value since Twister2 internally models the dataflow as a task graph and evenly distributes tasks over the cluster to balance the load. 

As shown in line 28, once the data transformation is performed, TSets can be converted to a different data format like NumPy such that it can feed to a different library to perform further processing. If Twister2 processes are bootstrapped using an MPI implementation, intermediate data can be directly processed using MPI API.

\begin{figure}[!h]
\begin{lstlisting}[language=Python, numbers=left,
    stepnumber=1,
    showstringspaces=false,
    tabsize=1,
    breaklines=true,
    breakatwhitespace=false]
from mpi4py import MPI
import numpy as np
from twister2 import Twister2Environment

env = Twister2Environment(resources=[{"cpu": 4, 
            "ram": 4096, "instances": 4}])
class PersonSource(SourceFunc):
    def next(self):
        # generate person tuple
        return ["id", "name", "address"]     
class VaccinationSource(SourceFunc):
    def next(self):
        # generate vaccination info. tuple
        return ["person_id", "doses"]   
def join_logic(student, result, ctx):
    return student["id"] == result["person_id"] 
                    and result["doses"] == 2 
        
people = env.create_source(PersonSource(), 
            parallelism = 10)
vaccination = env.create_source(VaccinationSource(), 
            parallelism = 10)
# finding people who have received two doses.
# this involves entire population, might 
# spill to the disk
fully_vaccinated = people.join(vaccination, 
            join_type=INNER, on=join_logic)
people_ids_split = fully_vaccinated.select("id").toNumpy()

# switching to use mpi directly
this_worker_total = people_ids_split.size
global_total = MPI.COMM_WORLD.allreduce(total, op=MPI.SUM)
people_ids = numpy.zeros(global_total, dtype=np.integer)
MPI.COMM_WORLD.allgather(people_ids_split, people_ids)                  
env.finalize() 	
\end{lstlisting}
\caption{Twister2 TSet on MPI}
\label{fig:tset_mpi}
\end{figure}

\subsubsection{Multidimensional Scaling}

\begin{figure}[h]
\centering
\includegraphics[width=0.25\textwidth]{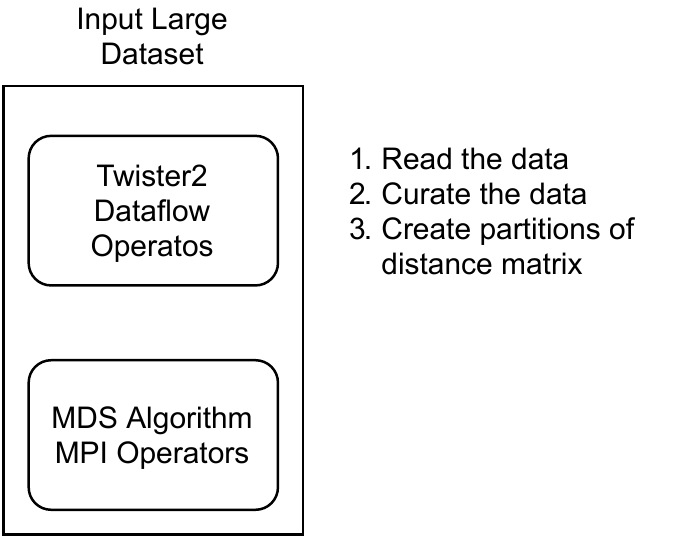}
\caption{Dataflow operators to prepossess data and MPI operators for Matrix manipulations in MDS algorithm.}
\label{fig:mds}
\end{figure}

\begin{figure}[h]
\centering
\includegraphics[width=0.35\textwidth]{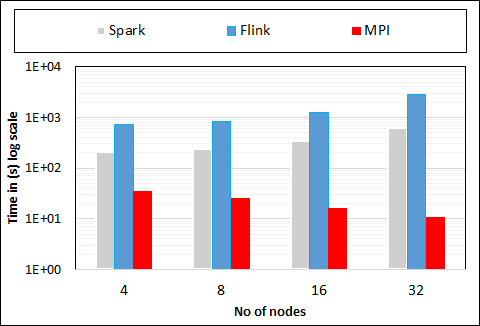}
\caption{MDS execution time with 32000 points on varying number of nodes. Each node runs 20 parallel tasks.}
\label{fig:mds_perf}
\end{figure}

Multidimensional scaling (MDS) is a valuable tool in data scientists toolbox. Authors have previously developed an MPI based MDS algorithm~\cite{spidal_java, java_threads, kamburugamuve2018anatomy} that can scale to large number of cores. The MDS algorithm expects a distance matrix and we need to calculate this matrix from an input dataset which can be large. This distance matrix is partitioned row-wise and feed into the algorithm. We developed an application that combines the MPI based MDS algorithm and Twister2 based data processing to create the partitioned distance matrix as shown in Figure~\ref{fig:mds}. This program runs executing table operators for data prepossessing and matrix operators for MDS algorithm. Further, Figure~\ref{fig:mds_perf} shows the strong scaling performance of MDS algorithm (only the algorithm) on varying number of nodes with 32000 points. Spark and Flink implementations of the MDS algorithm are developed on their table abstractions and MPI version is developed with the array abstractions and operators.

\subsection{Cylon}

Cylon~\cite{widanage2020high, abeykoon2020data} provides a distributed memory DataFrame API on Python for processing data using a tabular format. Unlike existing state-of-the-art data engineering tools written purely in Python, Cylon adopts high performance compute kernels in C++, with an in-memory table representation. Cylon uses the Apache Arrow memory specification for storing table data in the memory. It can be deployed with MPI for distributed memory computations processing large datasets in HPC clusters.

\begin{figure}[h]
\centering
\includegraphics[width=0.4\textwidth]{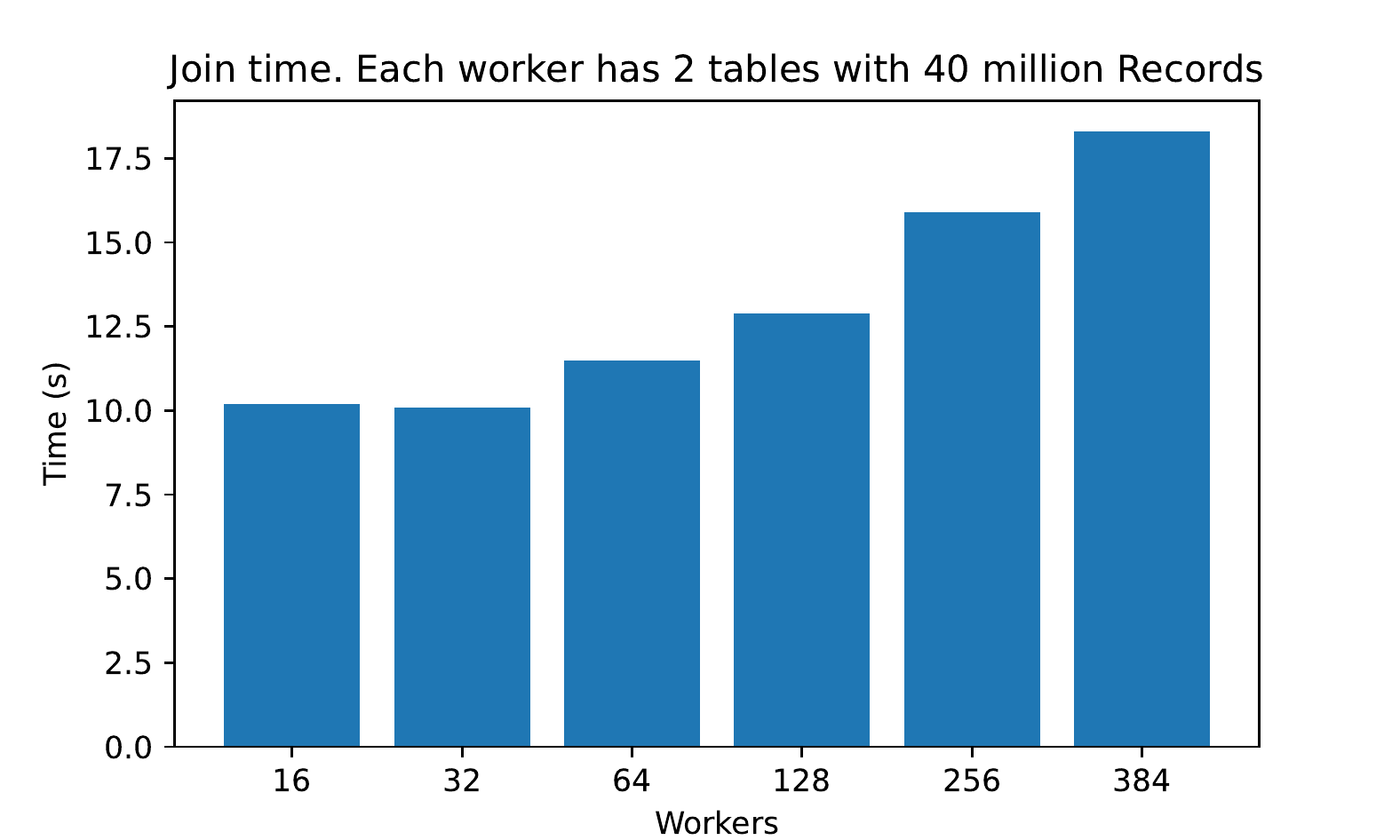}
\caption{Cylon join operator. Each worker is assigned two tables with 40 million records each. Each table has two 64bit integers columns.}
\label{fig:join_time}
\end{figure}

Operators in Cylon are based on relational algebra and closely resemble the operators in Pandas DataFrame to provide a uniform experience for the user. They are implemented as eager operators. The user can program with a global view of data by applying operations to them. Also, they can convert the data to local parallel processes and do in-memory operations as well. Cylon programs are SPMD-style programs and operators are designed to work in that fashion. Figure~\ref{fig:join_time} shows how Cylon Join operator can scale to large number of cores with increased load without sacrificing the performance. This test was done on a cluster with 8 nodes each with two Intel(R) Xeon(R) Platinum 8160 CPUs and 256GB of memory.

In Figure \ref{fig:cylon_mpi}, transformed data can directly pipe to a different framework like PyTorch~\cite{paszke2019pytorch} to train ML/DL models or extract more valuable insights from the data. Line 18 of Figure \ref{fig:cylon_mpi} transforms Cylon DataFrame into a NumPy array. Similarly, Cylon can convert DataFrames to multiple other data formats, including Pandas DataFrames, Arrow Tables, and even copy/move to a different device such as GPUs. Cylon tries to handle such transformations as efficiently as possible by avoiding additional data copies or wasting CPU cycles solely for data format transformation. Cylon owes this capability to its high performance core written in C++ and the columnar data representation it internally uses to hold data in memory. All these capabilities integrate Cylon operators seamlessly with existing frameworks and libraries without compromising the performance or adding additional memory pressure.

When Cylon processes are bootstrapped with an MPI implementation, the application gets all the capabilities of the underlying MPI implementation, including access to the BSP-style collective communication API. As shown in line 39, the programmer can use AllReduce operation of MPI to synchronize the model across the distributed set of workers. In addition to writing such custom code to handle synchronization, if the integrated library has the native capability to use MPI (PyTorch, Horovod~\cite{sergeev2018horovod}, TensorFlow~\cite{tensorflow}, Keras~\cite{geron2019hands} etc.), the programmer can use such capabilities since the process (Cylon Worker) already belongs to an MPI world. In addition, NVIDIA NCCL~\cite{jeaugey2017nccl} provides a BSP mode of execution for model synchronization at scale for GPU devices for accelerated deep learning. Functionality in NCCL is similar to MPI. Distributed interfaces like that of PyTorch have been specifically designed to provide a unified interface for accelerated deep learning on various accelerators. Additionally, Horovod also extends to GPU-based AllReduce for deep learning models. Thus Cylon can be seamlessly integrated not only for CPU-based MPI model synchronization, but GPU-based model synchronization as well. 



\begin{figure}[!h]
\begin{lstlisting}[language=Python, numbers=left,
    stepnumber=1,
    showstringspaces=false,
    tabsize=1,
    breaklines=true,
    breakatwhitespace=false]
import numpy as np
from mpi4py import MPI
from pycylon import DataFrame, read_csv, CylonEnv
from pycylon.net import MPIConfig

# preprocessing data using Cylon
env = CylonEnv(config=MPIConfig())
# load people
people = read_csv("people.csv", slice=True, env=env) # [id, severity]
# load vitals
vitals = read_csv("vitals.csv", slice=True, env=env) # [id, type, value]
# consider only temperature
temp_of_people = vitals.where(vitals["type"] == "TEMP")
people = people.set_index(["id"])
temp_of_people = temp_of_people.set_index(["id"])
# join temperature to people
joined = temp_of_people.join(people, how="left")
numpy_arr = joined.to_numpy()
# Create random input and output data
x = numpy_arr.T[1] # temperature
y = numpy_arr.T[3] # severity
# Randomly initialize weights
w = np.random.rand(4)

learning_rate = 1e-6
for t in range(2000):
    # Forward pass: compute predicted y
    # y = a + b x + c x^2 + d x^3
    y_pred = w[0] + w[1] * x + w[2] * x ** 2 + w[3] * x ** 3
    # Compute and print loss
    loss = np.square(y_pred - y).sum()
    # Backprop to compute gradients of a, b, c, d with respect to loss
    grad_y_pred = 2.0 * (y_pred - y)
    grads = mp.array([grad_y_pred.sum(), (grad_y_pred * x).sum(), 
            (grad_y_pred * x ** 2).sum(), (grad_y_pred * x ** 3).sum()])
    # Update weights
    w -= grads * learning_rate
    # synchronizing the model parameters using MPI
    w = np.array(MPI.COMM_WORLD.allreduce(w, op = MPI.SUM))/MPI.COMM_WORLD.Get_size()

env.finalize()
\end{lstlisting}
\caption{Cylon interoperability with Pytorch and MPI}
\label{fig:cylon_mpi}
\end{figure}

\subsection{Global Data \& Asynchronous Operators}

The data model in Apache Spark~\cite{spark2010} is based on Resilient Distributed Datasets (RDDs). RDD is an abstraction to represent a large dataset distributed over the cluster nodes. The logical execution model is expressed through a chain of transformations on RDDs by the user. A graph created by these transformations is termed the \textit{lineage graph}. It also plays an important role in supporting fault tolerance. 

Dask~\cite{rocklin2015dask} is similar to Spark in its execution and data model. The major difference is that Dask is implemented on top of Python as opposed to Java in Spark. Modin~\cite{petersohn2020towards} is a framework designed for scaling Pandas DataFrame-based applications. Modin allows Pandas DataFrame-based transformations to scale to many cores. 

Apache Spark, Dask and Modin all provide a global view of data. The user cannot access a parallel process and must program functions that are applied to the partitions of the global data. Also, they are executed with the asynchronous execution model we showed earlier. This makes these systems incompatible with the \archname{} architecture. 

\section{Conclusions}

This paper exploits the HPTMT principle that efficient distributed operators designed around a collection of key data abstractions enables an environment supporting high performance data-intensive applications at scale. We noted that many successful systems have used this principle but it is typically applied to a subset of available operators and data abstractions. We gave details of the HPTMT architecture for developing distributed operators independent of any framework that can work together using many orchestration (workflow) systems. We introduced the Cylon project, which provides eager operators on tables and the capability to use MPI's array (matrix)-based operators. This also provides an efficient bridge between different languages (C++, Python, Java) and links to the Java Twister2 project that provides DataFlow operators supporting this architecture. We believe that we have implemented enough operators across a range of application domains to show that one can achieve efficient parallel HPTMT across the thousands of operators found by aggregating scientific computing, Spark and Flink Big Data, NumPy, Pandas, Database, and Deep Learning. Our work focused on traditional CPU hardware but NVIDIA, Intel, and others are applying the HPTMT architecture to GPUs and accelerators. In future work, we will continue to explore more operators and their use across a variety of important hardware platforms. We will test these ideas on applications to verify end-to-end performance and that the operator-based programming model is indeed expressive enough. We can expect that we and others will discover the need for further operators as application experience develops. Apache Arrow and Parquet provide important tools for HPTMT and we are exchanging ideas with them. 

\section*{Acknowledgments}
This work is partially supported by the National Science Foundation (NSF) through awards CIF21 DIBBS 1443054, SciDatBench 2038007, CINES 1835598 and Global Pervasive Computational Epidemiology 1918626. We thank the FutureSystems team for their infrastructure support. 
\balance
\bibliographystyle{IEEEtran}   
\bibliography{ref.bib}

\end{document}